# Van der Waals heteroepitaxy of air stable quasi-free standing silicene layers on CVD epitaxial graphene/6H-SiC

*Zouhour BEN JABRA[1], Mathieu ABEL*[1], Filippo FABBRI[2], Jean-Noel AQUA[3], Mathieu KOUDIA[1], Adrien MICHON[4], Paola CASTRUCCI[5], Antoine RONDA[1], Holger VACH[6], Maurizio DE CRESCENZI[5], Isabelle BERBEZIER*[1]*

[1] Aix Marseille University, CNRS, IM2NP, Marseille 13397, France

[2] NEST, Istituto Nanoscienze – CNR, Scuola Normale Superiore, Piazza San Silvestro 12, 56127 Pisa, Italy

[3] Institut des Nanosciences de Paris, Sorbonne Université, CNRS, INSP, UMR 7588, 75005 Paris, France

[4] Université Côte d'Azur, CNRS, CRHEA, Valbonne 06560, France

[5] Dipartimento di Fisica, Università di Roma Tor Vergata, Roma 00133, Italy

[6] LPICM, CNRS, Ecole Polytechnique, IP Paris, Palaiseau 91128, France

**Corresponding Authors**

isabelle.berbezier@im2np.fr and mathieu.abel@im2np.fr

ABSTRACT

Graphene, consisting of an inert, thermally stable material with an atomically flat, dangling bond-free surface is by essence an ideal template layer for van der Waals heteroepitaxy of two-dimensional materials such as silicene. However, depending on the synthesis method and growth parameters, graphene (Gr) substrates could exhibit, on a single sample, various surface structures, thicknesses, defects, and step heights. These structures noticeably affect the growth mode of epitaxial layers, e.g. turning the layer-by-layer growth into the Volmer

Weber growth promoted by defect-assisted nucleation. In this work, the growth of silicon on chemical vapor deposited epitaxial Gr (1 ML Gr/1ML Gr buffer) on 6H-SiC(0001) substrate is investigated by a combination of atomic force microscopy (AFM), scanning tunneling microscopy (STM), x-ray photoelectron spectroscopy (XPS), scanning electron microscopy (SEM) and Raman spectroscopy measurements. It is shown that the perfect control of full-scale almost defect-free 1 ML Gr with a single surface structure and the ultra-clean conditions for molecular beam epitaxy (MBE) deposition of silicon represent key prerequisites for ensuring the growth of extended silicene sheets on epitaxial graphene. At low coverages, the deposition of Si produces large silicene sheets (some hundreds of nanometers large) attested by both AFM and SEM observations and the onset of a Raman peak at 560 cm$^{-1}$ very close to the theoretical value of 570 cm$^{-1}$ calculated for free-standing silicene. This vibrational mode at 560 cm$^{-1}$ represents the highest ever experimentally measured value and is representative of quasi-free standing silicene with almost no interaction with inert non-metal substrates. From a coverage rate of 1ML, the silicene sheets disappear at the expense of 3D Si dendritic islands whose density, size, and thickness increase with the deposited thickness. From this coverage, the Raman mode assigned to quasi-free standing silicene totally vanishes, and the 2D flakes of silicene are no longer observed by AFM. The experimental results are in very good agreement with the results of kinetic Monte-Carlo simulations that rationalize the initial flake growth in solid-state dewetting conditions, followed by the growth of ridges surrounding and covering the 2D flakes. A full description of the growth mechanism is given. This study, which covers a wide range of growth parameters, challenges recent results stating the impossibility to grow silicene on a carbon inert surface and is very promising for large scale silicene growth. It definitely shows that silicene growth can be achieved using perfectly controlled and ultra-clean deposition conditions and an almost defect-free Gr substrate.

INTRODUCTION

Silicene (Si-ene) represents an ideal 2D material that shares all the outstanding characteristics of graphene (mainly magnetic, mechanical, optical and electrical properties)[1-7] associated with the additional advantage of a narrow band gap that can easily be opened by doping,[8-11] by applying external electric and magnetic fields,[12, 13] and mechanical strain.[14] It also has the significant potential for straightforward integration into industrial semiconductor production lines, making Si-ene a promising candidate for microelectronic devices.[15-19] Si-ene has therefore attracted a large amount of studies and has been the subject of intense debates about the origin of the complex surface superstructures observed as well as its strong interaction with the substrate.[20-24]

The first evidence of monolayer silicon with anomalous optical properties (optical absorbance at 280 nm and photoluminescence peak at 434 nm, i.e. 2.9 eV), was given using chemical exfoliation from calcium disilicide with propylamine hydrochloride[25, 26]. However, it was then reported that unlike Graphene (Gr), Si-ene cannot be obtained by mechanical exfoliation.[27] Si-ene should then be deposited by molecular beam epitaxy (MBE)[28] or chemical vapor deposition (CVD).[29]

The studies on Si-ene initially focused on its growth on Ag(001),[30] Ag(110),[31] and Ag(111) substrates.[20, 32, 33] The growth of graphene-like Si-ene nanoribbons was first reported. The demonstration of larger Si-ene sheets was then obtained by a combination of scanning tunneling microscopy (STM), angular-resolved photoemission spectroscopy (ARPES), and density functional theory (DFT) simulations.[34, 35] However, rapidly various controversies have emerged about the origin of the various surface reconstructions observed, including (4×4), (√13×√13)R13.9°, (√7×√7)R19.1° and (2√3×2√3)R30° with respect to Ag(111).[23, 36-38]

First-principles calculations demonstrated that the linear dispersion band observed by ARPES experiments could originate from the hybridization states between silicene and Ag(111) surface.[39] Besides, its impressive stability was definitely ascribed to covalent bonds with the substrate.[21, 40] It was also reported that the Dirac cone in Si-ene on a Ag(111) surface was destroyed due to the distortion of the Si-ene crystalline cell and strong band hybridization at the Si-ene/Ag(111) interface by unfolding bands.[41] The linear dispersion observed in ARPES was then ascribed to the Ag substrate, instead of Si-ene.[39]

Consequently, the investigations naturally shifted to alternative metal substrates such as Ir(111),[42] Ru (0001),[43] Pt(111),[44] Pb(111),[45] and Au(111).[46, 47] There again, strong interactions with the substrates were demonstrated both theoretically and experimentally.[48, 49]

More recently, to overcome this limitation various experiments have been carried out on C-based inert substrates such as highly oriented pyrolytic graphite (HOPG) with hexagonal atomic arrangements.[50-53] On this substrate, the results demonstrated that the charge modulations caused by quantum interferences (QI) serve as templates and guide the incoming Si atoms to self-assemble in 2D clusters.[54] The restricted extension of these QI limits the growth to very tiny 2D Si-ene flakes in the nanometer range and causes the transition to a 3D growth mechanism resulting in small 3D clusters at the outskirts of the tiny 2D Si-ene areas as observed experimentally.[54, 55] The size limitation also originates from the presence of high defect densities on the HOPG substrate which favor both intercalation of silicon underneath the top surface (even at room temperature) resulting in the formation of buried silicene [56, 57] and the preferential defect-assisted 3D cluster nucleation at the origin of the transition between layer-by-layer growth and Volmer-Weber growth.

The van der Waals (vdW) epitaxy of Si-ene layers with low-buckled honeycomb geometry on graphene-covered SiC substrates was theoretically demonstrated by first-principles calculations including vdW interaction in the ground state and quasiparticle effects in the electronic structure.[58] However, similar behavior (intercalation and 3D cluster nucleation) was also observed during the growth of silicon on graphene epitaxially grown on Ruthenium[59, 60] or on Ir substrates.[61] At the opposite, a defect-free atomically planar graphene surface, can act as scaffold to stabilize the growth of 2D materials, enabling construction of 2D materials that do not naturally exist or are not stable as free-standing films.[62]

To go beyond and meet the challenges raised by microelectronic devices, semiconducting $MoS_2$ [63, 64] and insulating h-BN[65] substrates were used for the vdW epitaxy of Si-ene. However, considering the complexity of obtaining $MoS_2$ with high quality buffer layers (monocrystalline and free of defects and moirés) and h-BN over large areas without crystal defects or impurities, the use of such stacked layers seems cumbersome to date for any potential applications.[66] In summary, since most of the inert substrates have sizeable density of defects, a general tendency of Si atoms to intercalate during deposition or thermal annealing has been evidenced.

In this work we have investigated the growth of silicon deposited in ultra-clean Molecular Beam Epitaxy (MBE) conditions on an almost defect-free graphene layer obtained on 6H-SiC(0001) substrate by an original CVD process already described elsewhere.[67, 68] We show that two prerequisites are essential for the epitaxial growth of large scale Si-ene flakes on Gr: first, perfectly flat almost defect-free 1ML Gr and second the ultra-clean UHV environment during the growth of Si-ene. By a combination of atomic force microscopy (AFM), STM, x-ray photoelectron spectroscopy (XPS), scanning electron microscopy (SEM) and Raman spectroscopy measurements, we demonstrate the formation of 2D quasi-free standing Si-ene flakes on a scale of one-two hundred nanometers randomly deposited all over the sample surface. In absence of surface defects and contaminants both on the substrate and in the deposited material, both the diffusion path toward the subsurface layer and the preferential nucleation of 3D islands are significantly reduced. Without interaction with the Gr substrate, silicon forms 2D quasi-freestanding flakes whose lateral extension reaches some hundreds of nanometers. The growth mechanism of 2D Si-ene flakes is explained by kinetic Monte-Carlo (KMC) simulations that perfectly reproduce and explain the growth morphologies resulting from 2D growth in solid-state dewetting conditions followed by the growth of 3D ridges surrounding the Si-ene flakes.

RESULTS

Figure1a shows a STM image of the nominal Gr film on 6H-SiC(0001) (after cleaning at 650°C), where the Raman measurements in Figure 1d were performed. The absence of defects was observed in numerous large-scale AFM and SEM images recorded on each sample (see for instance Figures S1). The morphology presents terraces separated by a train of steps with 1.5 nm height, which mimics the regular train of 6 silicon carbide bilayer steps of the 6H-SiC(0001) substrate. The steps are oriented along $[1\bar{1}00]$ and are constituted of smaller sections with armchair edges, forming an angle of 120° to each other.[68] Figure 1b gives a STM high resolution image of the surface structure where the superimposition of the graphene crystalline cell (0.246 nm) and the $(6\sqrt{3} \times 6\sqrt{3})R30°$ superstructure referring to $6\sqrt{3}$xSiC cell or 13xGr cell (~ 3.2nm) is clearly visible. Such a surface is called (6x6)Gr (6xSiC cell = 1.9nm) in the rest of the manuscript (see Ref. [69] for a detailed description of graphene ). The corrugation of the top surface is about 0.02 – 0.04 nm in good agreement with the literature.[70] A complete description of the surface structure and composition of (6x6)Gr can be found in Ref. [71]. The sample is fully covered by very stable 1ML (6x6)Gr which is free of contaminants and of extended defects thanks to the cleaning procedure. The presence of point defects cannot be excluded. The XPS peak of carbon (C1s) gives evidence for the absence of oxygen at the surface (Figure 1c). It can be deconvoluted into 4 different components ascribed to the buffer layer (BL) S1 (284.6eV) and S2 (285.3 eV), to the Gr on top of the BL (284.2 eV) and to SiC below the BL (283.4 eV) that are representative of Gr on BL on 6H-SiC(0001).[72,73] The quantification of the Gr coverage has been previously estimated to 1ML (from XPS analysis of the Si/C ratio).[68]

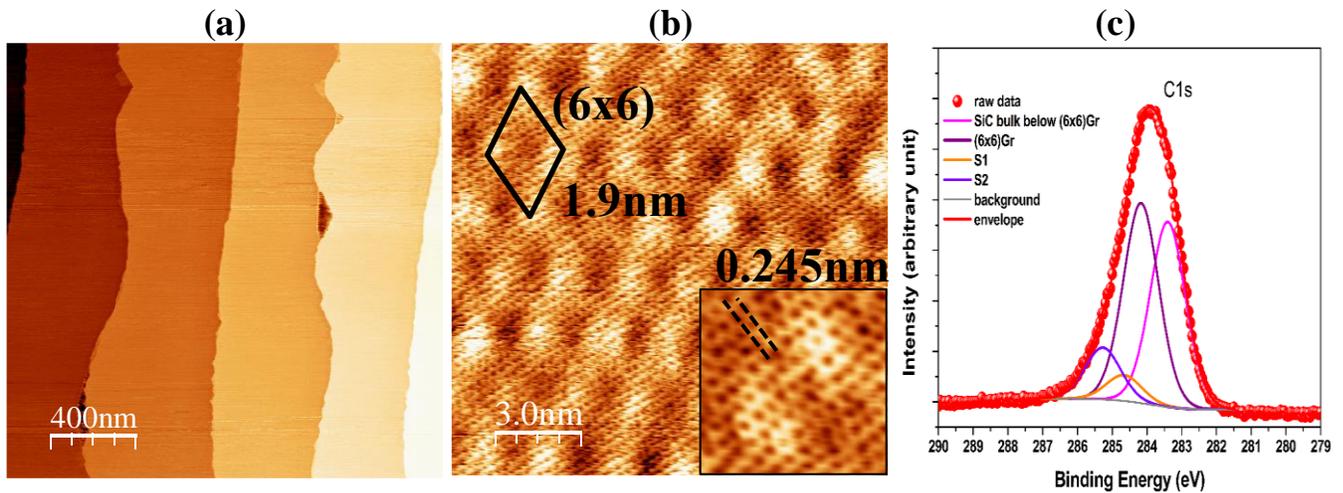

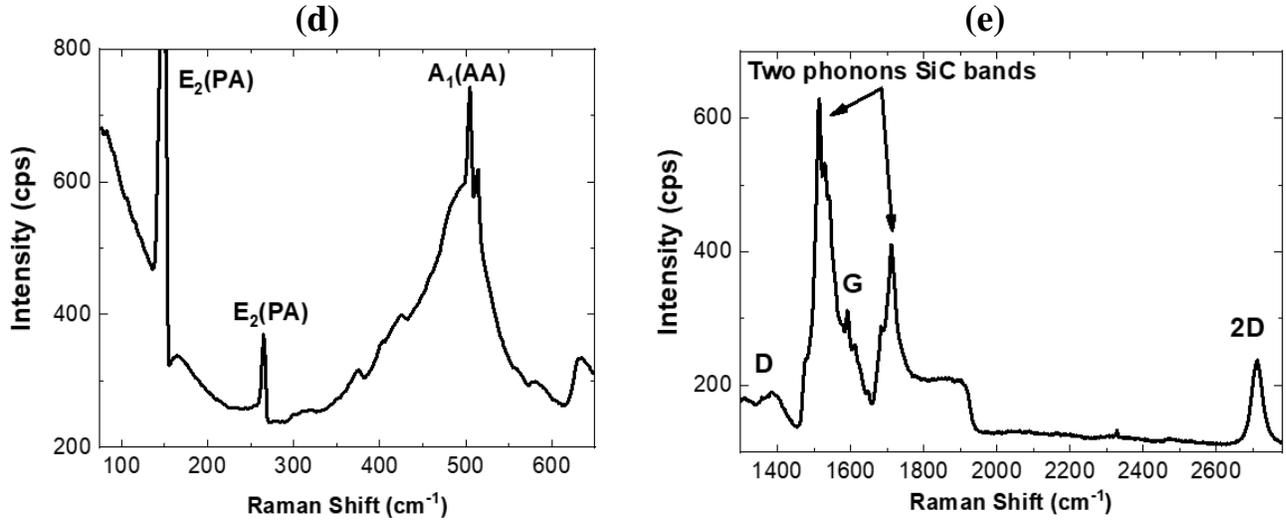

*Figure 1: Characterization of the nominal Gr surface after thermal cleaning at 650°C: (a) and (b) STM images of the Gr surface at low and high magnification respectively; (c) XPS spectrum; Raman spectra in the silicon (d) and in the graphene range (e) where the D, G and 2D peaks can be observed together with Raman modes due to the 6H-SiC(0001) substrate.*

The Raman spectrum of the epitaxial graphene in the silicon range is shown in Figure 1d, presenting the different Raman features of the 6H-SiC(0001) substrate. In particular, in such a range the different Raman peaks are attributed to acoustic modes: the acoustic planar E2 mode set at 149 cm$^{-1}$ and at 265 cm$^{-1}$, the axial acoustic A1 doublet set at 505 cm$^{-1}$ and 514 cm$^{-1}$.[74] In addition two small peaks at 380cm$^{-1}$ and 430cm$^{-1}$ can be ascribed to nitrogen doping of the SiC substrate.[75] Raman spectrum in the high Raman shift range, Figure 1e, shows the three fingerprints of the electron-phonon interaction in 2D Gr, i.e. the D peak (at 1358cm$^{-1}$), the G peak (1591cm$^{-1}$) and the 2D peak (2712cm$^{-1}$). The D peak is ascribed to defects associated to step edge orientation in addition to other defects (atomic vacancies, dislocations, grain boundaries) of the substrate and its intensity is related to the concentration of such defects.[76,77] . In the present case, the D/G intensity ratio is rather low. From a careful deconvolution of the peaks of the Figure 1e, it amounts to 0.09, corresponding to a defect density of 1.9x10$^{10}$ cm$^{-2}$,[78] i.e. to about one defect over 10$^5$ atoms of the graphene surface value well lower than the ratio obtained in exfoliated Gr where it reaches values up to 0.8.[79] In addition, since the D peak onset requires intervalley scattering, it appears near a perfect armchair edge (while it cannot be produced by a perfect zigzag step edge). This could also explain the presence of the D peak in the (6x6)Gr samples studied here that have armchair step edges (as already demonstrated [68]). Such behavior does not apply to the 2D peak, not involving intervalley scattering, and that can be activated both by armchair and zigzag edges. The 2D peak is the second order overtone of the D peak. The graphene spectrum has a 2D peak with a single Lorentzian shape and with a full width at half-maximum FWHM(2D) ~35 cm$^{-1}$, a signature of single layer graphene on a SiC substrate.[80] This result confirms the already reported self-limited graphene growth on top of the BL/6H-SiC(0001) which produces 1ML Gr.[68] The G peak corresponds to the doubly degenerate Raman active optical vibration E2g mode, where the carbon atoms oscillate in the graphene plan. The 6H-SiC(0001) exhibits various overtone peaks between 1000 to 2000 cm$^{-1}$ with some of them superimposed to the peaks of Gr, i.e. the peak at 1514 cm$^{-1}$ which is the overtone of the TO(X) phonon at 761 cm$^{-1}$ and the peak at 1713 cm$^{-1}$ which is a combination of optical phonons with wave vectors near the M point at the zone edges.

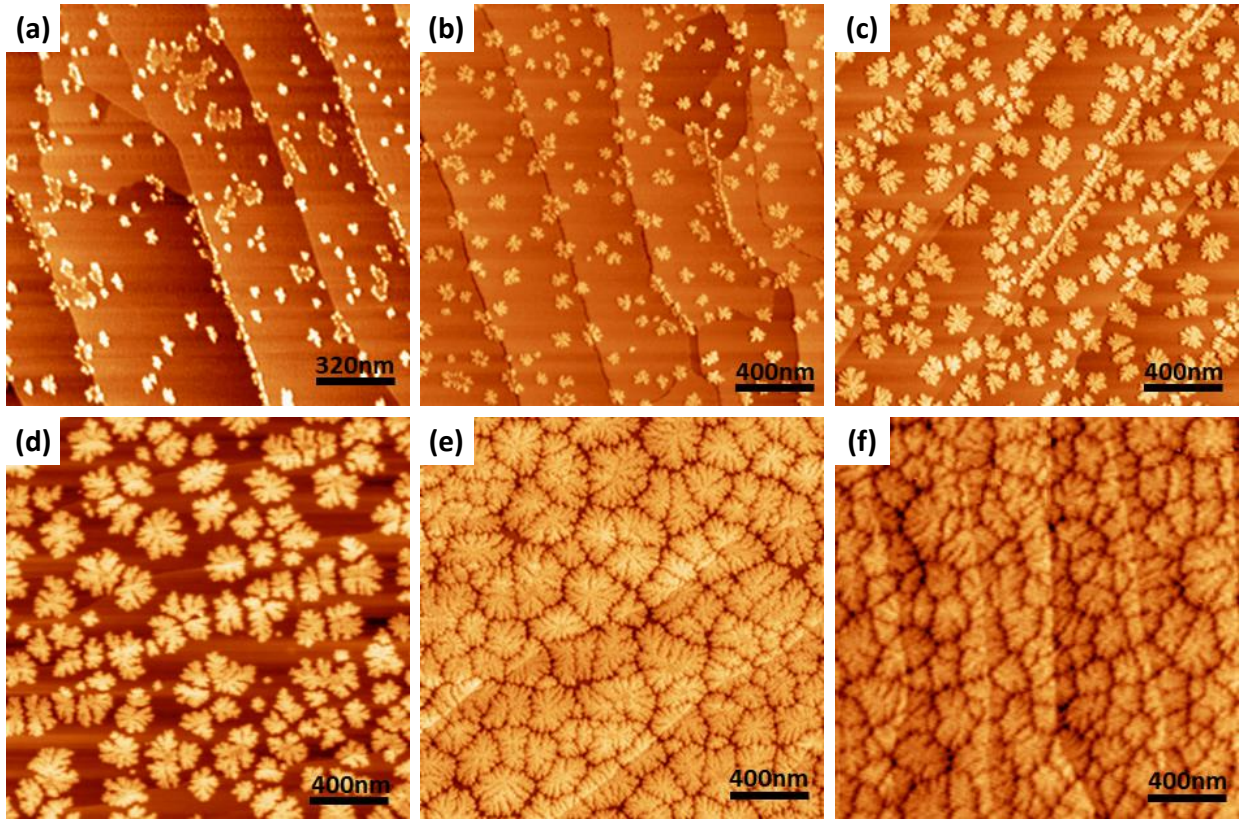

*Figure 2: AFM images of the surface after MBE deposition at room temperature on (6x6)Gr/6H-SiC(0001) substrate of increasing thickness θ of silicon : (a) 0.125ML; (b) 0.25ML; (c) 0.5ML; (d) 1MLs; (e) 3MLs; and (f) 6MLs.*

During the deposition (at room temperature) of silicon layers of increasing thickness from θ = 0.125ML to 6MLs at a deposition rate of 0.2 nm/min, the morphology of the surfaces changes entirely (Figure 2). The deposition of 0.125 ML results in small 3D islands and large flat domains, randomly and sparsely distributed over the surface. The former have an average diameter of 100 nm ±10 nm. A more detailed observation (zoom of images in Figures S2 and S3) shows that the larger domains consist of two parts: a flat central area with a thickness of ~ 0.35nm ± 0.02nm and lateral dimensions about one-two hundreds of nanometers, surrounded by small ridges with a mean height of 1.2nm ± 0.2nm. At increasing thickness (0.25 and 0.5ML), the size of the ridge surrounding the flat areas progressively increases (both laterally and in height) and their dendritic shape becomes more pronounced. The 2D flat central areas keep constant their lateral dimensions and thickness up to a 0.5ML coverage, at which they start to be less visible, due to the presence of the ridges (i.e. higher dendritic islands on their edges). Since the density of the flakes increases with the deposited thickness, the overall amount of 2D material increases up to 0.5ML. After 1 ML deposition (Figure 2d), the 2D flat areas totally disappear in favor of 3D dendritic islands and the coverage rate of the surface by dendrites is about 47%. From this 1 ML coverage rate, only the growth of 3D dendrites is observed, resulting in an increase in dendrites size and height with the deposited thickness. After 3ML deposition (Figure 2e), the surface is fully covered by dendrites, whose only thickness increases as the amount of the deposit increases up to 6MLs (Figure 2f). Scanning electron microscopy has been used to confirm the homogeneity of the deposition over the samples and their morphology (Figure S2). STM characterization of the Si deposition was very complicated due to contamination of the STM tip by Si clusters. In a few rare cases we obtained images like the ones reported in Figure S4 of the Supplementary Materials, showing the presence of darker and brighter regions. Despite the noise on the large scale STM images (Figures S4c and S4d), it is well visible that both areas are characterized by the superimposition of the two (6x6)Gr and the Gr lattices (Figures S4e and S4f). Darker regions have a dendritic shape, resembling the shadow of the dendritic islands observed by AFM (see Figure 2).

The elemental and chemical composition of the surface is characterized by XPS. Focus is put on Si2p and C1s peaks that are recorded at increasing deposited thicknesses of Si (Figure 3). First, it can be noted that the C1s peak remains constant throughout the growth process and regardless of the deposited thickness (Figure 3a), with the position and shape of (6x6)Gr already described above and with a similar deconvolution into 4 components (S1, S2, SiC bulk and top layer Gr). The absence of any disturbance (shift or shoulder) of this peak is a first indication of a very weak interaction of the Gr substrate with the Si coating. On the other hand, most of the Si2p peak (~ 101.3eV) observed comes from the SiC substrate. At the coverage of 3MLs, the appearance of two shoulders is well visible. They are ascribed to a SiOx peak at 102.3eV and the Si bulk at 99.2eV. The Si2p peak evolution has been studied in detail in the Supplementary Materials (see Figures S5). This analysis shows that the Si and $SiO_x$ peaks (around 99-99.5 eV and 102.6- 103.4eV respectively) start to be detectable for Si coverage about 0.5ML where in parallel, the dendritic ridge becomes well visible. Probably the air resistance of the silicence sheets is due to the passivation of the flake edges by these dendritic structures. Due to their 3D silicon crystallinity the ridges act as getter for the oxygen, having a preferential oxidation of such structures. The areas of the $SiO_x$ and Si bulk peaks increase with the Si coverage becoming more and more visible for the 3 and 6 ML. At these coverages, the intensity ratio between the SiC substrate peaks and the Si and SiO$_x$ peaks decreases hugely due to the formation of a quasi-continuous layer of 3D dendrites oxidized during air exposure. According to literature,[81, 82] the Si 2D-peaks on inert substrates are expected around 99-99.5 eV, i.e. at very close energy to the clean Si 3D-peaks. So, it is not straightforward to distinguish the 2D from the 3D clean Si 2p XPS peaks. Therefore, we suggest that the signal located around 99.5 ± 0.3 eV can be ascribed to both the Si-ene flakes and the 3D dendrites. This is consistent with the AFM analysis which evidences, even for the lowest Si coverages, the presence of Si-ene flakes surrounded by ~1nm thick Si ridges.

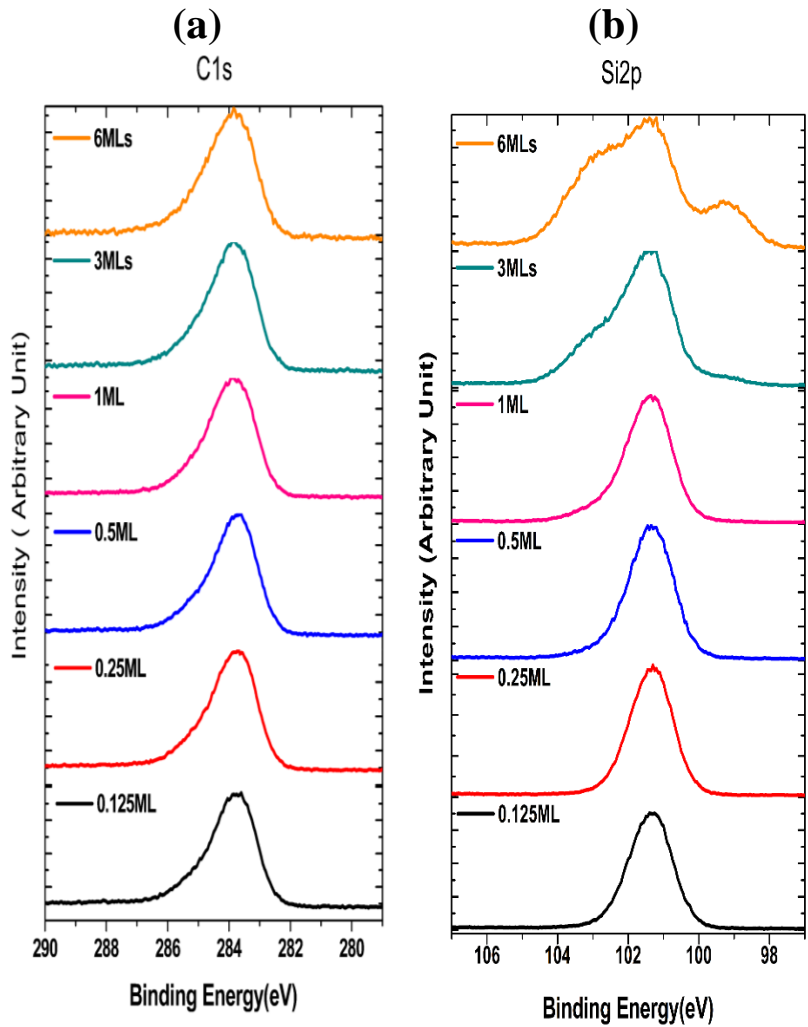

*Figure 3: XPS spectra of Si thin films of different thicknesses from 0,125, to 0,25, 0,5, 1, 3, and 6MLs: (a) C1s and (b) Si2p core-level peaks.*

Figure 4a displays the Raman spectra for Si layers with different coverages grown on the (6x6)Gr layer. The first-order asymmetric peak located at 520 cm$^{-1}$ can be assigned to the 6H-SiC(0001) substrate. For small coverages (< 1 ML) two new peaks appeared: one at 560 cm$^{-1}$ which is interpreted as the zone-center E2g vibrational mode of Si-ene (equivalent to the so-called G-like peak of Gr), which was predicted by theoretical studies at 570 cm$^{-1}$ [34] and another at 240 cm$^{-1}$ which is assigned to the breathing mode (a zoom of the two peaks is given in Figure 4a). A breathing mode at 230 cm$^{-1}$ was already reported for Si-ene on Ag. It was correlated with the out-of-plane displacement of Si-atoms that is generated by the buckling.[83] Its intensity is affected by the quantity and nature (zigzag or armchair) of step edges in the Si-ene flakes and we can expect that a higher density of 2D Si-ene flakes produces a peak with higher intensity. In this series of experiments, its intensity slightly increases up to 0.5ML until it completely disappears at 1ML, as shown in Figure 4b. A similar evolution is observed for the E2g peak: it increases with coverage rate up to 0.5ML where it reaches its maximum, before it disappears at 1ML coverage. The absence of any silicon related Raman modes at high coverage (>1ML) suggests the oxidation of the silicon deposited on the surface. The two peaks are absent for larger coverages, meaning that Si-ene is only formed at small deposited thicknesses (<1ML). These results are in good agreement with the AFM images of the surfaces (Figure 2) where 2D flakes are well visible until 0.5ML coverage, from which on they totally disappear. We can then conclude that the 2D flakes visible on the surfaces are composed of 2D quasi-free-standing Si-ene.

The increase of the E2G peak intensity until 0.5ML (Figure 4b) is most likely induced by an increase of the quantity of Si-ene (well visible on Figures S2 and S3). Nevertheless, with such low coverage rates (0.125, 0.25 and 0.5) and quite inhomogeneous areas distribution, it is very uncertain to give a quantitative estimate of the Si-ene coverage on the surface.

It is the first time that the E2g peak of Si-ene is observed at a frequency so close to the free-standing peak confirming the very low interaction of the Si-ene flakes with the epitaxial graphene. In addition, as it has been already reported, the E2g mode in 2D materials can also be modulated by strain and doping.[34] In the case of Si-ene on Ag, the E2g peak is located at 530cm$^{-1}$ because of its strong interaction with the substrate which is obviously not the situation here.[84] In the case of Si-ene nanosheets on HOPG substrate, the E2g mode is located around 542 cm$^{-1}$ because of the relaxation of the Si-Si bond length due to removal of the periodic constrains for an infinite Si-ene layer with respect to the surface underneath . [51]

In addition, the Raman shifts of the E2g and breathing mode are reported in Figure 4c. It is worth noting that the breathing mode Raman shift remains at 240 cm$^{-1}$ for different silicon thicknesses, supporting the attribution to the out-of-plane displacement of silicon atoms[83], while the E2g Raman shift increases linearly with the silicon coverage from 558 cm$^{-1}$ (0.125 ML) up to 563 cm$^{-1}$ (0.5 ML). This is an indication that, with the enlargement of the lateral size of the Si-ene flakes, their interaction with the substrate decreases leading the $E_{2g}$ Raman shifts towards the freestanding value of 570 cm$^{-1}$. It could also be explained by the presence of oxidized Si islands at the periphery of the domains which could affect the strain and doping of the Si-ene flake. The error bars reported in Figure 4b and 4c are obtained by statistical analysis of 20 μm x 20 μm Raman maps.

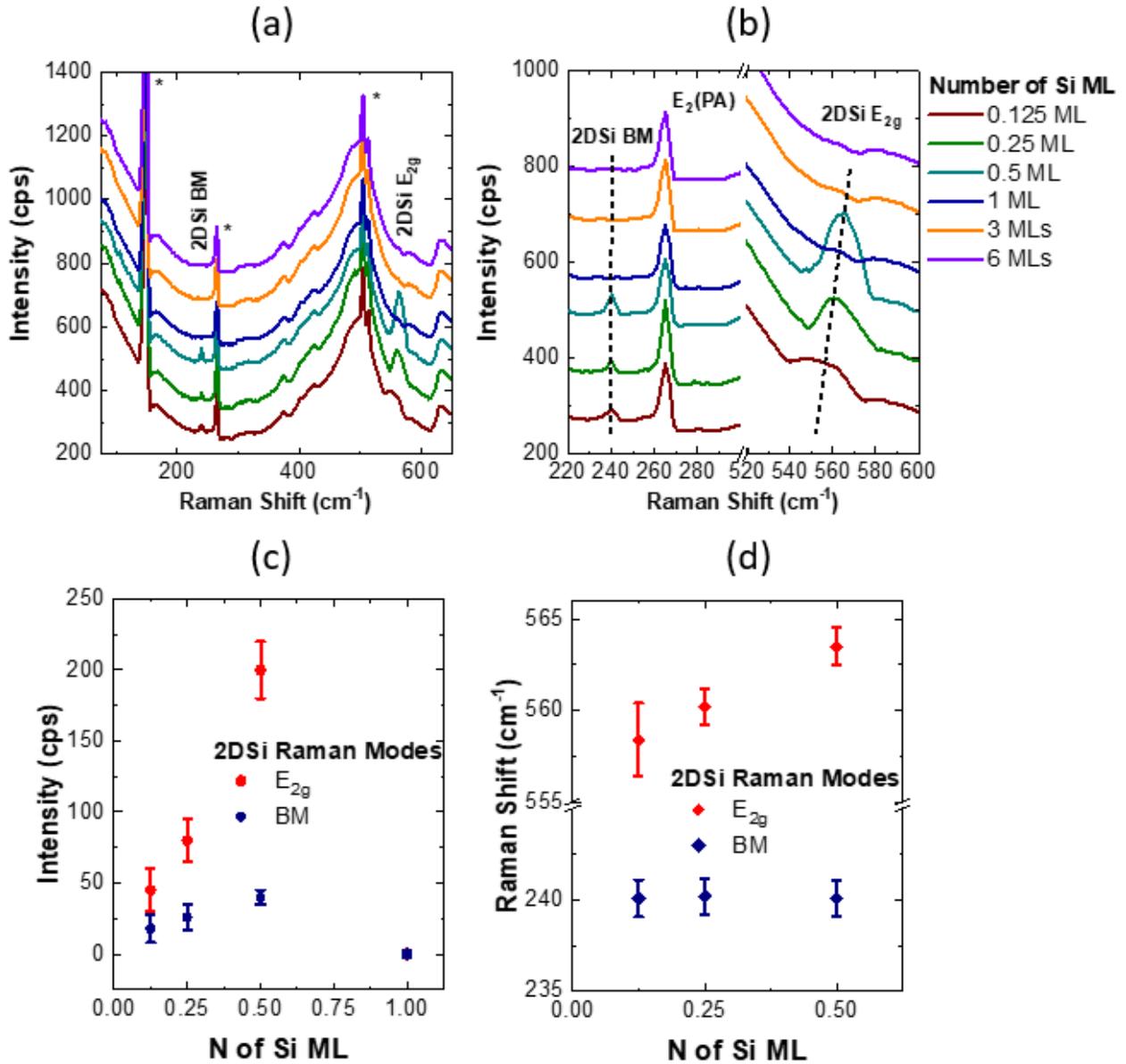

*Figure 4: Raman spectra of Si/(6x6)Gr/6H-SiC(0001) with 0,125, 0,25, 0,5, 1, 3 and 6ML coverage: (a) silicon vibrational range of the major Raman features of silicene. Asterisks denote the SiC Raman modes. Evolutions of the 2D Si-ene mode (b) intensity and (c) shift.*

The evolution of the Raman spectrum of epitaxial graphene with increasing coverage of silicon is shown in Figure 5a. The main modifications of the graphene Raman modes are: the broadening of the 2D mode, as summarized in Figure 5b, and the shift of the G and 2D modes. In particular, the full width at half maximum (FWHM) of the 2D mode increases with the increase of the silicon coverage, going from 35 cm$^{-1}$ for the 0.125 ML up to 38 cm$^{-1}$. The error bars reported in Figure 5b are obtained by statistical analysis of Raman maps obtained in a 20 μm x 20 μm sampling area. The broadening of the 2D peak is assigned to nanoscale strain effects on the graphene layers.[85] Considering the low interaction of the Si-ene sheets obtained on the graphene surface, it is possible that the strain effects are induced by the presence of the dendritic structures, and in fact this effect dominates for the 3 ML and 6 ML coverages, where Si-ene is completely absent and a complete coverage with a dendritic layer is observed. A similar behavior is reported in the correlation plot of the Raman shift of the G peak with the one of the 2D mode (Figure 5c). The dispersions with different silicon coverages reveal that the n-type doping of epitaxial graphene slightly decreases, while an important spreading is present in all the set of data along the strain direction. This decrease of the electron doping concentration can be due to interface states of oxidized silicon at high silicon coverage.[86] The slope of the linear fitting for the different dispersions varies between 1.2 up to 1.7 that is an indication for biaxial strain of the graphene layer, which is expected due to the complex interaction between graphene and the dendritic structures.[87]

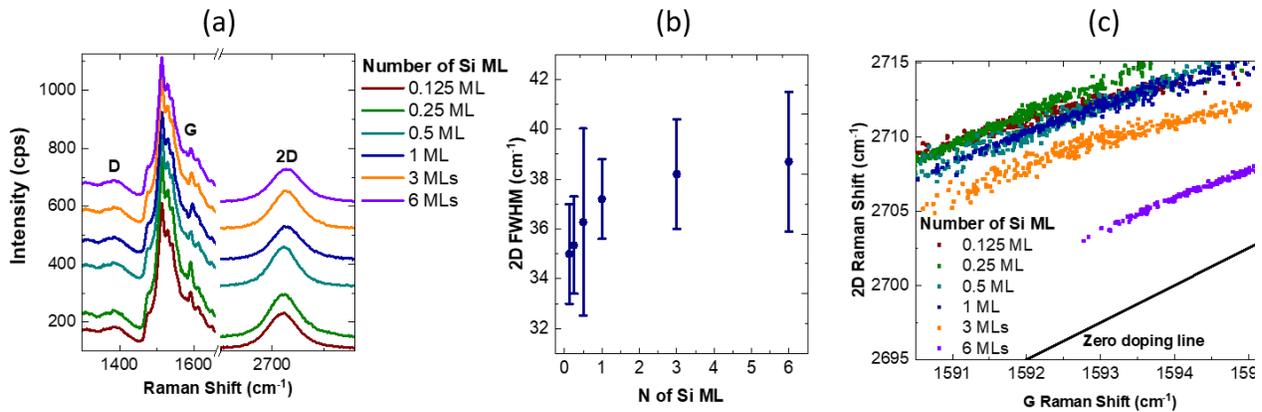

*Figure 5: Raman spectra of Si/(6x6)Gr/6H-SiC(0001) with 0,125, 0,25, 0,5, 1, 3 and 6MLs coverages: (a) vibrational range of the major Raman features of graphene. (b) Evolution of the FWHM of the 2D mode as function of the silicon coverage. (c) Strain/doping correlation plot of epitaxial graphene with increasing silicon coverage.*

We have then annealed Si-ene samples to better understand the mechanism of formation and stabilization of thin Si-ene areas surrounded by 3D dendrites. Annealings were performed (both in situ just after Si deposition and in another UHV setup after the samples have been exposed to air) for samples with a deposited thickness of 0.5 ML since the most intense Raman peaks resulting from the most extended 2D Si-ene areas are obtained at this coverage. After deposition, all samples showed evidence for large flakes of Si-ene (with lateral sizes up to one-two hundreds of nm) with compact almost hexagonal shapes surrounded by dendritic 3D islands (Figure 6a). On some AFM images, it is also well visible that the Si-ene "floats" on top of the Gr substrate without interacting with it, as can be seen on inset of Figure 6a, where a flake is straddling two terraces like a sheet laid on the step edge. A typical example of the morphology obtained after 2h annealing at 850°C is given on Figure 6b., The density, size and shape of the 2D flakes do not evolve with annealing, but their edges become straighter following directions at 120° from each other, which further emphasizes their pseudo-hexagonal shape representative of higher quality crystalline flakes (Figure 6b). The height of the compact planar areas remains in the range of 0.2 – 0.3nm.

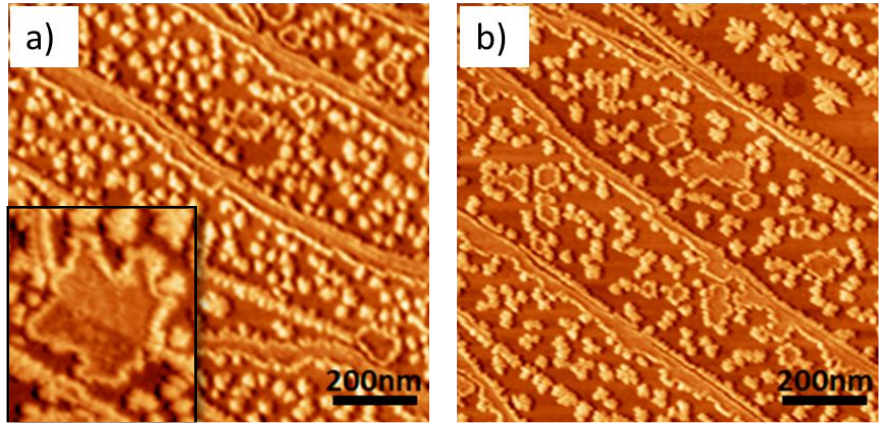

*Figure 6: AFM images of the surface after 0.5ML Si deposition at RT: (a) just after growth; in the inset is presented the image of a Si-ene flake straddling two terraces; (b) after 2h annealing at 850°C.The corresponding STM images of this sample are presented in figure S4.*

The initial steps of the Si-ene flake growth (2D domains followed by 2D domains with a ridge) may be perfectly rationalized by an epitaxial growth mode dictated by solid-state dewetting.[88] We show evidence for such a mechanism thanks to kinetic Monte-Carlo simulations of the epitaxial dynamics. We consider a model geometry with a Solid On Solid (SOS) model where the atomic columns are defined on the basis of the Gr honeycomb lattice. A constant deposition flux $F$ allows the deposition of Si on Gr. Meanwhile, the top-most atoms are allowed to diffuse with a diffusion energy barrier $E_S$ and to detach from their $n$ nearest-in-plane neighbors with a bonding energy barrier $E_N$. In this case, the diffusion and detachment processes occur with a frequency proportional to $\exp[-\beta(E_S + n\,E_N)]$. Importantly, the energy barriers $E_S$ and $E_N$ are supposed to be function of the local height $h$. This is characteristic of wetting conditions where excess quantities depend on the local environment, and especially on the thickness here. This crucial assumption happens naturally here because the Si atoms on the first layer are only weakly bounded to the Gr layer via van-der-Waals interactions. On the contrary, Si atoms diffusing on top of already deposited Si atoms are expected to experience a greater energy barrier due to their covalent bonding with underlying Si atoms. These barriers are thence a priori function of the atomic height, until they reach their values on bulk Si beyond a given thickness (typically beyond 3 or 4 monolayers).[89] The diffusion barrier of Si on Gr was computed to be as low as 0.06 eV on the basis of ab-initio calculations.[90] It is indeed much lower than the usual diffusion barrier of Si on Si(001) that was measured to be 0.7 eV.[91] The system is highly anisotropic and the dependence of $E_N$ on $h$ is less documented. However, for the special case of Si on top of Si(001) with anisotropic type-A and B terraces, it was typically estimated at 0.5 eV by comparison of the results of such a SOS model with anisotropic experimental morphologies on Si(001).[92] We thence considered that $E_S$ and $E_N$ vary over $h=1$ and 2 and stay constant for $h \geq 3$. We fixed $E_S$ ($h=1$) = 0.06 eV and $E_S$ ($h \geq 3$) = 0.7 eV and considered typically $E_S$ ($h=2$) =0.08 eV, while we set $E_N$ ($h$) = 0.4, 0.5 and 0.6 eV for $h$ =1,2 and 3, respectively. The resulting evolution is displayed in Figure 7.

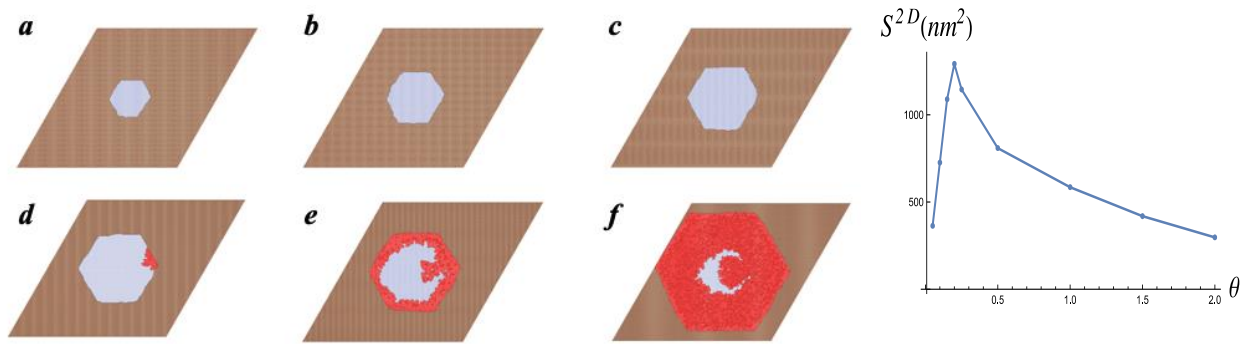

*Figure 7: (left) Morphologies of KMC simulations of a SOS-model with thickness-dependent diffusion $E_S$ and binding energies $E_N$ at room temperature for the typical parameters described in the text and for coverage Si amount $\theta$ = (a) 0.05, (b) 0.1, (c) 0.15, (d) 0.2, (e) 0.5 and (f) 2.0 ML (the system size is 200 x 200 graphene lattice parameters, brown region). (right) Evolution of the surface of the 2D flakes as a function of $\theta$.*

Initially, growth proceeds by the enlargement of a flake that remains perfectly flat and 2D (Figure 7a-c). From a typical coverage of $\theta^*$, and for flake sizes of the order of 100 lattice parameters, a ridge starts to form on one edge of the flake. It is noticeable that this localized ridge perfectly corresponds to the geometries initially observed experimentally (Figure 2a). Once formed, the ridge quickly spreads over the entire flake edge and then thickens slowly. At $\theta$=0.5ML (Figure 7e), the 2D zone with a thickness $h$=1ML is entirely surrounded by a ridge that has a typical thickness of 2 to 3 ML, once again in accordance with the flakes observed experimentally. The precise values of $\theta$ where the latter steps occur depend on statistical fluctuations, but we observe that the ridge is usually fully formed at 0.5ML. As deposition proceeds, the 2D flake is gradually shrinking and the ridge progressively occupies the entire flake, as can be evidenced by the evolution of the 2D central zone in Figure 7e-f. We thence argue that the so particular flakes and ridges morphologies observed experimentally result from the epitaxial growth dynamics of 2D systems with dewetting conditions depicted by our model. The values of the energy barriers considered here are therefore adequate to account for the experiments, and this result can be considered as a validation or at least an indication of their real values. Important variations of these barriers could, indeed, lead to other growth modes: a plane-by-plane growth; a fast dewetting leading to compact rough and thick shapes, etc. A detailed study of the different morphologies potentially contained by this model will be given elsewhere. Another parameter of the model concerns the initial nucleation of the 2D flakes. We considered here a perfectly homogeneous nucleation on a flat surface without defect. This corresponds to 2D flakes located on the terraces. For the flakes that overflow from one terrace to another and cross an atomic step, we can intuit that their nucleation was initiated at the step. While this geometry does not fundamentally change the mechanism of a de-wetting growth dynamic, it may affect the densities of islands which will be the subject of a future study.

The above modeling is dedicated to the 2D flake and ridge formation and does not capture the dendritic growth that would require a full description of the Si aggregate growth. The mechanism of transition from 2D to 3D dendrites in MBE growth is poorly understood, even if in general it is correlated to the growth rate and deposition flux. Here the situation is very complex since Si is supposed to have a very high diffusion length on Gr even at RT,[90] and there is a huge anisotropy of interaction between the strong Si-Si lateral bonds and the weak interaction perpendicular to the surface. When the 2D compact domains reach a critical size, the rate at which the adatoms arrive at the domain edges becomes higher than the edge diffusion that becomes insufficient to ensure the rearrangement of all incoming adatoms to a stable geometrical configuration. This promotes the branching along the directions of the domain expansion, leading to dendritic Si domains. For this reason, the compact area will ramify into branched structures whose primary branches roughly follow a pseudo-hexagonal shape. For larger dendritic structures, with elongated branches, the adatoms agglomerate following a diffusion limited process which produces a thickening and disorientation of the branches and the 3D growth prevails. The structural transition can then be attributed to the difference between the rate of adatom attachment to the edge of the growing nuclei, the rate of edge diffusion and to the strong anisotropy of the Si-Si interactions between the lateral and the vertical ones. At increasing thickness, the morphology evolves with a densification and thickening of the 3D dendrites accompanied by an increase of their size. A comprehensive study on the growth mechanism of the 3D silicon dendrites will be the topic of a future work.

CONCLUSION

Our results provide a clear proof and a complete explanation of the van der Waals epitaxy of 2D Si-ene flakes on (6x6)Gr on 6H-SiC(0001), when using ultra-clean MBE deposition technique and almost defect-free graphene substrate. The 2D Si-ene flakes are well identified by AFM images for low silicon deposition thickness (<1ML). The presence of 2D flat areas straddling two terraces like a sheet laid on the step, gives a first indication that the Si-ene flakes "float" on top of the Gr substrate without interacting with it. The 2D domains are surrounded and progressively covered by 3D ridges forming dendritic islands. At 1ML deposition they entirely cover the surface at the expense of the 2D flakes. The formation of large scale Si-ene flakes is also unambiguously attested by Raman spectra which evidence the vibrational modes ascribed to Si-ene: the zone-center E2g vibrational mode at 560 cm$^{-1}$, which was predicted by theoretical studies at 563-570 cm$^{-1}$ (depending on the doping) for free-standing Si-ene and the breathing mode at 240 cm$^{-1}$ which is correlated to the out-of-plane displacement of Si-atoms generated by the buckling. The intensity of these two peaks, which is related to the quantity of Si-ene flakes, increases up to 0.5ML where the peaks reach their maximum until they completely disappear at 1ML. The absence of these two peaks for larger coverages, indicates that Si-ene areas are only present on the surface for small deposited thicknesses (<1ML).

The growth of the Si-ene flakes followed by 2D domains with a surrounding ridge is well explained by Kinetic Monte-Carlo simulations when considering highly anisotropic barriers i.e. a factor 10 difference between the diffusion energy barrier ($E_S$) and bonding energy barrier ($E_N$). These barriers are also function of the atomic height, until they reach their values on bulk Si beyond a critical thickness (3-4ML). The first layer of Si atoms which is only weakly bounded to the Gr layer via van-der-Waals interactions has a very low diffusion barrier (set at 0.06eV on the basis of ab-initio calculations). On the contrary, Si atoms diffusing on top of already deposited Si atoms are expected to have a greater energy barrier due to their covalent bonding with the underlying Si atoms (0.7eV was measured on Si(001)). A similar height dependent evolution of $E_N$ is also considered. In these conditions, it is shown that the growth proceeds by the enlargement of a flake that remains perfectly flat and 2D up to a typical deposited thickness, above which a ridge starts to form on the edge of the flake. Once the ridge is formed, it quickly spreads over the entire flake and then thickens slowly. The model demonstrates that the so particular flakes and ridges morphologies observed experimentally result from the epitaxial growth dynamics of 2D systems with dewetting conditions.

Our results elucidate the mechanism of formation of Si-ene and provide an efficient and simple way to produce high-quality and large-scale material on an inert and well-ordered surface. The work also enables investigation of quantum phenomena, topological insulator behavior towards reliable integration in devices with huge potential applications in nano-electronics, optoelectronics, and photonics.

## METHODS

Graphene samples were grown on the Si-terminated face of Silicon Carbide (6H-SiC(0001)) in a horizontal hot wall CVD reactor. The 6H-SiC(0001) substrates are on-axis n-doped (nitrogen) 2'' wafers with typical thickness of ~350 µm from Tankeblue. Typical residual offcuts are between 0.05 and 0.2° toward the $[1\bar{1}00]$ direction, as deduced from AFM measurements on annealed substrates under $H_2$.[93] Before introduction into the growth chamber, the substrates are cut into 1×1cm² pieces and then chemically cleaned with isopropanol. Graphene deposition was carried out in a horizontal CVD reactor allowing a homogeneous and reproducible deposition of Gr on the substrates, at a pressure of 800 mbar using a gas mixture of propane ($C_3H_8$) as the carbon source, and hydrogen ($H_2$) and argon (Ar) as the carrier gases. The $H_2$ flow ratio ($H_2$ flow / total flow) was set to 9% which ensures a surface totally covered by 1ML Gr on top of the buffer layer with the $(6\sqrt{3} \times 6\sqrt{3})R30°$ reconstruction (i.e. partly bonded to the SiC substrate).

The samples completely covered with the 1ML epitaxial Gr are introduced into a perfectly clean ultra-high-vacuum (UHV) molecular beam epitaxy (MBE) growth chamber dedicated to Si-based heterostructures. Various set of Si depositions were performed on two different types of epi-Gr grown on hexagonal SiC(0001) substrates. The substrates are characterized by two different density of defects according to Raman results (i.e. with different ratios between the intensity of the D and G graphene modes). On these two different substrates, the Si atoms deposited at room temperature on the epi-Gr give rise to different behaviors. In the present study, with D/G intensity ratio is around 0.09, i.e. high graphene quality, we obtained wide Si-ene regions, while in the case of D/G intensity ratio, reaching values up to 0.25, the Si atoms tend to penetrate the epi-Gr layer and form Si-ene intercalated nanosheets.

In addition, we also tested the effect of the Gr substrate cleanliness: when the substrate does not undergo any thermal cleaning, the silicon atoms form a rough amorphous layer of silicon without any Si-ene flakes (Figure S6).

The Gr samples are then first thermally cleaned at 650°C for 120 min. Then the heating is switched off and the samples are cooled down to room temperature during 180 min before silicon deposition. Silicon is deposited on the samples at room temperature using an electron beam evaporator, maintaining a constant deposition rate with uniformity of 5% on the whole sample (with sample rotation). The silicon flux was set to 0.2nm/min, which gives the best homogeneity of the deposit. The deposition time was varied between 0.2 and 1.6 min which corresponds to deposited thicknesses (θ) from 0.125ML to 6MLs.

The samples prepared in the above-mentioned way were thoroughly characterized by micro-Raman spectroscopy, performed using a Renishaw Invia Qotor equipped with a confocal optical microscope in back reflection geometry. A 532 nm excitation laser was employed with an excitation power of 0.5 mW and a 1800 line per millimeter grating (spectral resolution 2 cm$^{-1}$). The laser beam was focused onto the sample by a x100 objective with a numerical aperture NA=0.85 and a spot size of 800 nm. All the Raman spectra are acquired with an acquisition time of 3 s. All the Raman analyses were carried out in air after air exposure of the samples.

Structural characterization of the samples before and after silicon deposition was investigated by STM measurements in UHV conditions using electrochemically etched tungsten tips. The UHV chamber is equipped with low energy electron diffraction (LEED), monochromatic XPS and a STM (Omicron). STM observations were carried out at room temperature using constant current mode with It = 0.3 nA and Vt between 0.2 and 2.0 V. They were performed on the samples as grown and after annealing in UHV at temperatures around 850°C to assess the stability of the deposit. A photon energy of 1486.6 eV (Al-Kα) for monochromatized XPS measurements was used for all samples. The photoelectrons are analyzed with an Omicron EA 125 energy analyzer with a 30eV pass energy. The quantification through XPS analysis of the Si/C ratio, estimated a Gr coverage of 1ML as detailed elsewhere.[68] AFM images were obtained in non-contact mode in air at room temperature (Park Systems) using standard silicon cantilevers. AFM and STM data were analyzed by WSxM[94] and ImageJ software. The XPS spectra were fitted using the CasaXPS software. All the XPS and STM analyses were carried out in a UHV chamber different from the MBE growth one. Therefore, all the samples experienced air exposure during the transfer.

SUPPORTING INFORMATION

S-1. Large scale SEM and AFM images of the (6x6)Gr substrate

S-2. SEM Analysis

S-3. AFM additional images

S-4. STM images

S-5. XPS spectra of Si2p for different Si deposition on Gr/SiC

S-6. AFM images of the silicon surface after deposition on uncleaned Gr substrate

AUTHOR INFORMATION

**Corresponding Authors**

* isabelle.berbezier@im2np.fr and mathieu.abel@im2np.fr

**Author Contributions**

The manuscript was written through contributions of all authors. All authors have given approval to the final version of the manuscript. All the authors contributed equally.

**Funding Sources**

P.C. and M. D.C., would like to acknowledge the European Community for the HORIZON2020 MSC-RISE Project DiSeTCom (GA 823728).

HV would like to acknowledge the HPC centers of IDRIS (Grant A0100900642 ) and CERMM for computational resources.

**Acknowledgments**

The authors acknowledge the Nanotecmat platform. H.V. would like to acknowledge the HPC centers of IDRIS (Grant A0100900642 ) and CERMM for computational resources.